\documentclass[prl,preprint,groupedaddress,showpacs]{revtex4}
\usepackage{graphicx}
\begin{document}
\title{PARAMETRIC POTENTIAL DETERMINATION BY THE CANONICAL FUNCTION METHOD}

\author{C. Tannous}
\affiliation{Laboratoire de Magn\'{e}tisme de Bretagne, UPRES A CNRS 6135,
Universit\'{e} de Bretagne Occidentale, BP: 809 Brest CEDEX, 29285 FRANCE}

\author{K. Fakhreddine}
\affiliation{CNRS, BP: 11-8281, Beirut, Lebanon}

\author{J. Langlois}
\affiliation{Laboratoire des Collisions Electroniques et Atomiques,
Universit\'{e} de Bretagne Occidentale, BP: 809 Brest CEDEX, 29285 FRANCE}

\date{\today}

\begin{abstract}
The canonical function method (CFM) is a powerful means for solving the Radial Schrödinger Equation. The 
mathematical difficulty of the RSE lies in the fact it is a singular boundary value problem. The CFM turns it into a 
regular initial value problem and allows the full determination of the spectrum of the Schrödinger operator without 
calculating the eigenfunctions.Following the parametrisation suggested by Klapisch and Green, Sellin and Zachor we 
develop a CFM to optimise the potential parameters in order to reproduce the experimental Quantum Defect results for 
various Rydberg series of He, Ne and Ar as evaluated from Moore's data.
\end{abstract}


\maketitle

\section{INTRODUCTION}
A reliable calculation of the electron impact ionisation cross-section of an atom requires an accurate 
determination of the continuum wavefunctions in the incident and in the exit channels. At intermediate 
and low-energy it is necessary to have a fairly good representation of the interaction between the incoming 
electron and the target in the incident channel and between the outgoing electron and the residual ion in 
the exit channel. This work is concerned with the latter problem. \\

The main difficulty in the calculation of the final state continuum wavefunction arises from the non-local 
character of the exchange potential. Various alternatives have been used to overcome this difficulty, one 
of the most popular being the local exchange approximation of Furness and Mc Carthy ~\cite{Furness}. While this 
approximation seems to work well when the continuum energy is high enough ~\cite{Rash} it breaks down at low 
energy. This is precisely the case where our model potentials are expected to be the most accurate.\\

Our goal is to determine potentials which incorporate the effect of exchange while keeping a local 
character. The potential felt by an electron in the field of an ion is not exactly Coulombic at intermediate 
distances. The difference between the effective and Coulomb potentials manifests itself through the 
Quantum Defects (QD) associated with bound states and the  phaseshifts associated with continuum 
states. For a given angular momentum l, the QD $\sigma_l$ is defined by:

\begin{equation}
E_{nl}=-\frac{Z^2}{{(n- \sigma_{l})}^2}
\end{equation}

where $ E_{nl}$ is the binding energy of the excited electron in the nl orbital and Z the charge of the ionic core.
As n increases the QD is constant for a given Rydberg series. An interesting account of the early and 
recent development of this concept has been given by Rau and Inokuti ~\cite{Rau}. It should be noted that when 
the ionic core is an open shell  system, there is in principle a different value of the QD for each coupling 
scheme between the core and the excited electron spin and angular momenta.\\

When the electron energy is above the continuum limit, its wavefunction is characterised by a phaseshift 
for each spin and angular momentum. The difference between the actual and the Coulombic phaseshifts $\delta_l$ 
is the signature of the non-Coulombic part of the electron-ion interaction. Near the continuum limit it is 
related to the QD through:

\begin{equation}
\delta_{l}=\pi \sigma_l
\end{equation}

Thus, our approach is to determine a parametric potential  from the information contained in the bound 
state spectrum of the electron-ion system. That potential might be used to calculate the continuum 
wavefunctions.

An extensive review of the applications of model potentials has been given by Hibbert ~\cite{Hibbert}. Two 
particularly interesting parametric potentials will be considered here. The first one is due to Klapisch ~\cite{Klapisch}:

\begin{equation}
V(r) = -(2/r) [ 1+(Z-1) exp(-\alpha_{1} r)+ C r \hspace{0.2 cm} exp(-\alpha_{2} r) ]
\end{equation}

where  $\alpha_1$, $\alpha_2$ and C are adjustable parameters. This type of potential has been widely used by Aymar et al. 
(see Aymar ~\cite{Aymar} and references therein). The second has been suggested by Green, Sellin and Zachor (GSZ) ~\cite{Green}

\begin{equation}
V(r)= -(2/r) [(Z-1) \Omega(r)+1]
\end{equation}

where:

\begin{equation}
\Omega(r)=1/( \epsilon_1 [exp(\frac{r}{\epsilon_2})-1]+ 1 )
\end{equation}

where $\epsilon_1$ and $\epsilon_2$ are adjustable parameters. Both expressions are Coulombic near the origin and at large 
distances.\\

The canonical function method is a powerful means for calculating the QD for a given potential since it  
provides the eigenspectrum of the Schrödinger operator accurately and quickly without having to 
determine the eigenfunctions.

\section{THE CANONICAL FUNCTION METHOD}

The canonical function method (CFM) ~\cite{Kobeissi 82} is a powerful means for solving the Radial Schrödinger 
Equation (RSE). The mathematical difficulty of the RSE lies in the fact it is a singular boundary value 
problem. The CFM turns it into a regular initial value problem and allows the full determination of the 
spectrum of the Schrödinger operator bypassing the evaluation of the eigenfunctions.\\

The Canonical Function Method (CFM) developed by Kobeissi and his coworkers ~\cite{Kobeissi 90, Kobeissi 91} to integrate the 
RSE consists of writing the general solution as a function of the radial distance r in terms of two basis 
functions $\alpha(r)$ and $\beta(r)$. Picking an arbitrary point $r_0$ at which a well defined set of initial conditions are 
chosen ie: $\alpha(r_0)=1$ with $\alpha'(r_0)=0$ and $\beta(r_0)=0$ with $\beta'(r_0)=1$, the RSE is solved by progressing towards the 
origin and towards Infinity. During the integration, the ratio of the functions is monitored until saturation 
signaling the stability of the eigenvalue spectrum.\\

Several advantages to the CFM are worth mentioning. First, the RSE integration belonging to the Singular 
Boundary Value class is avoided by transforming the problem into an initial value one with well-defined 
starting values for the basis functions $\alpha(r)$ and $\beta(r)$. Second, the RSE eigenvalue problem is solved and 
the eigenfunctions are simultaneously determined once the roots of the eigenvalue function (EF) are 
found. The EF is defined as the difference of the ratio $\frac{\alpha(r)}{\beta(r)}$ at saturation for large and small r ~\cite{Kobeissi 82}. The EF, depicted in Figure 1 below, behaves typically as a tan function when the spectrum has stabilised after 
the saturation of the basis function ratio with r.\\

The speed and accuracy of the method have been tested and compared to standard integration algorithms 
in a variety of cases and for a wide of range of potentials. In addition the method has been generalised to 
the phase shift estimation as well to the continuum coupled channel problem  ~\cite{Kobeissi 91} ~\cite{Kobeissi 90}.\\

In potential optimisation problems the method can only be of great value because of its speed, accuracy 
and ease of programming. Accordingly, one has to pick an appropriate optimisation method and use the 
CFM as a subroutine for a given parametric potential.\\

\begin{figure}[htbp]
\begin{center}
\scalebox{0.9}{\includegraphics*{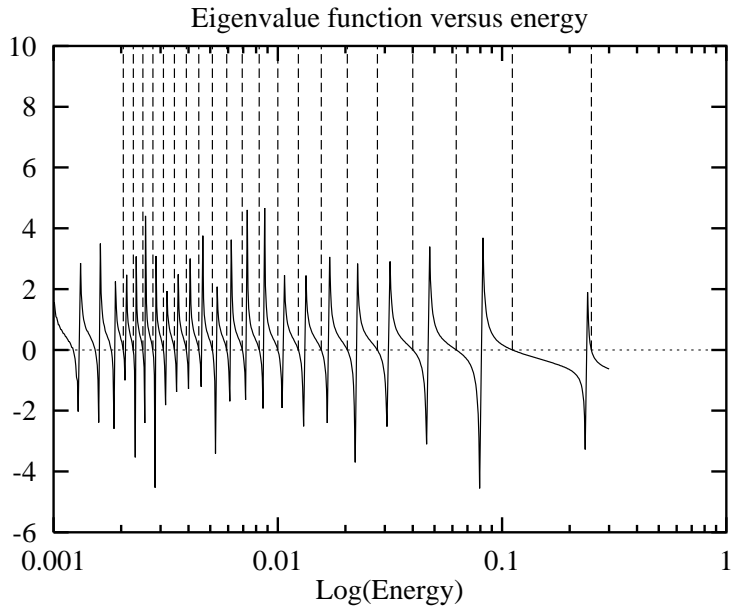}}
\end{center}
\caption{Behavior of the eigenvalue function with energy. The vertical lines indicate the eigenvalue position.}
\label{fig1}
\end{figure} 

We choose the well documented Klapisch ~\cite{Klapisch} and Green et al. ~\cite{Green} 
parametric potentials and Helium, Neon and Argon as tesbeds. The experimental values that we 
start from are the levels and  quantum defects constituting a stringent test for the validity
of the optimised potentials.\\

\section{RESULTS}

In table 1, we display the Helium (L=0) singlet and triplet results in the Klapisch (entry with $\alpha_1$, 
$\alpha_2$ and C) and GSZ (entry with $\epsilon_1$ and $\epsilon_2$) cases along with the corresponding QD. The CFM result is  indicated with QD whereas the experimental value is indicated by Exp.

For Argon states (l=0, l=1 and l=2) the optimisation based on the CFM gives with Klapisch and GSZ 
potentials are displayed in table 2.

For Neon (S=0 with l=0, l=1 and l=2) states using the GSZ potential, we find with the CFM the results
displayed in table 3.

\section{DISCUSSION}

From the above tables a remarkable agreement between the experimental and optimised QD is observed 
in most cases. While a slight agreement might be observed in some instances, one should not attribute it 
to the CFM or to the optimisation procedure but on the large uncertainties or limitations observed in the 
experimental values. The tabulated QD are averaged over the full spectrum that might contain 20 or 30 
levels obtained from the CFM, while the Moore data did not provide more than 10 levels in the same case. 
Judging from the agreement with the level and QD values in some cases, the Klapisch potential 
description might be superior to the GSZ, however it does poorly in the Neon case. We believe the GSZ 
potential yield with our formulation based on the CFM, results that appear more adequate to Neon than 
the Klapisch potential. Nevertheless we cannot rule out entirely the Klapisch description simply because 
its optimisation parameter space is larger and perhaps our optimisation procedure did not sample 
uniformly the entire space. Our work related to optimisation of Klapisch potential for Neon is still in 
progress.

{\bf Acknowledgements}:

This paper is dedicated to the memory of Dr. Hafez Kobeissi (1936-1998). Helpful correspondance, 
regarding mathematical aspects of the RSE, with Dr. Jeff Cash are gratefully acknowledged. Dr. Stéphane 
Mazevet provided preliminary results for the triple differential cross section versus slow electron angle in 
e-2e experiments etablishing further the validity of our optimised potentials.

\newpage

\appendix
\begin{center}
{\bf TABLES}
\end{center}

\begin{table}[h]
\begin{center}
\begin{tabular}{|l||c|c|c|r|}
\hline

S=0  & $\alpha_1$= 6.32313 & QD= 0.14& $\epsilon_1$=0.556962& QD=0.21 \\
     & $\alpha_2$ =4.65907 & (Exp)= 0.13 & $\epsilon_2$= 0.529966  & (Exp)= 0.13 \\
     & C= 5.69105  & & & \\
\hline
\hline
 
S=1  & $\alpha_1$= 2.00428 & QD= 0.30& $\epsilon_1$= 0.411661& QD=0.29 \\
     & $\alpha_2$= 0.998444 & (Exp)= 0.30 & $\epsilon_2$=0.688626 & (Exp)= 0.30 \\
     & C= 0.654791 & & & \\
\hline     
\end{tabular}
\caption{Parametric potentials for Helium (Klapisch and GSZ for l=0, S=0 and S=1)}
\end{center}
\label{tab1}
\end{table}

\begin{table}[h]
\begin{center}
\begin{tabular}{|l||c|c|c|r|}
\hline

l=0  & $\alpha_1$= 9.87780E-02 & QD= 2.51& $\epsilon_1$= 0.115144& QD= 2.16\\
     & $\alpha_2$ = 1.00056E-01& (Exp)= 2.16 & $\epsilon_2$= 0.105113  & (Exp)= 2.16 \\
     & C= 7.90032E-02  & & & \\
\hline
\hline

l=1  & $\alpha_1$= 1.07863 & QD= 1.64& $\epsilon_1$= 0.999785& QD=1.64 \\
     & $\alpha_2$ = 1.02077& (Exp)= 1.73 & $\epsilon_2$= 0.981995  & (Exp)= 1.74 \\
     & C= 1.01160  & & & \\
\hline
\hline

l=2 & $\alpha_1$= 0.856959 & QD= 0.30& $\epsilon_1$= 0.117336 & QD=0.32 \\
     & $\alpha_2$ = 1.08286 & (Exp)= 0.36 & $\epsilon_2$= 0.499883  & (Exp)= 0.36 \\
     & C= 10.39858  & & & \\

\hline     
\end{tabular}
\caption{Parametric potentials for Argon (Klapisch and GSZ for l=0, l=1 and l=2)}
\end{center}
\label{tab2}
\end{table}

\begin{table}[h]
\begin{center}
\begin{tabular}{|l||c|r|}
\hline

l=0  & $\epsilon_1$= 28.2369 & QD=1.32 \\
     & $\epsilon_2$= 15.7071  & (Exp)= 1.32 \\

\hline
\hline

l=1 & $\epsilon_1$= 5.54229 & QD=0.88 \\
     & $\epsilon_2$= 1.47477  & (Exp)= 0.89 \\

\hline
\hline

l=2 & $\epsilon_1$= 2.96440 & QD=0.046 \\
     & $\epsilon_2$= 0.873570  & (Exp)= 0.032 \\

\hline 
    
\end{tabular}
\caption{Parametric potentials for Neon (GSZ for l=0, l=1 and l=2)}
\end{center}
\label{tab3}
\end{table}

\end{document}